\documentclass[usegraphicx,usenatbib,usedcolumn,useAMS]{mn2e}

\newcommand{\per}[1][1]{{\scriptsize$^{-#1}$}}
\newcommand{\kms}{km~s\per}
\newcommand{\unisim}{\sim\!}
\newcommand{\asymerr}[2]{{\tiny\makebox[0pt][l]{\raisebox{-0.75ex}{$-#1$}}%
        \raisebox{+1.25ex}{\makebox[0pt][l]{$+$}\phantom{$-$}$#2$}}}
\newcommand{\slantfrac}[2]{#1\!\left/#2\right.}

\newcommand{\vrot}{V_\rmn{rot}}
\newcommand{\rdspec}{r_\rmn{d,spec}}
\newcommand{\rdphot}{r_\rmn{d,phot}}
\newcommand{\reff}{r_\rmn{eff}}
\newcommand{\mueffapp}{\mu_{R,\rmn{eff},\rmn{app}}}
\newcommand{\mueffabs}{\mu_{R,\rmn{eff},\rmn{abs}}}
\newcommand{\dtf}{\Delta {M_B}^{\!\rmn{TF}}}
\newcommand{\snr}{$\slantfrac{S}{N}$}
\newcommand{\paperone}{Paper~I}
\newcommand{\elfitpy}{\textsc{elfit2py}}
\newcommand{\elfitd}{\textsc{elfit2d}}
\newcommand{\gimd}{\textsc{gim2d}}
\newcommand{\sextractor}{\textsc{SExtractor}}
\newcommand{\eqnref}[1]{Eqn.~\ref{#1}}
\newcommand{\vltthanks}{\thanks{
    Based on observations made with ESO Telescopes at
    Paranal Observatory under programme IDs 066.A-0376 and 069.A-0312.}}
\newcommand{\hstthanks}{\thanks{
      Based on observations made with the NASA/ESA Hubble 
      Space Telescope, obtained from the data archive at the Space 
      Telescope Institute. STScI is operated by the association of 
      Universities for Research in Astronomy, Inc. under the NASA 
      contract NAS 5-26555.}}

\title[The TFR of distant field galaxies]%
      {The Tully--Fisher relation of distant field galaxies%
        \vltthanks\hstthanks}
\author[S. P. Bamford et al.]{
  S. P. Bamford$^{1}$\thanks{E-mail: ppxspb@nottingham.ac.uk},
  A. Arag\'on-Salamanca$^{1}$ and B. Milvang-Jensen$^{2}$\\
  $^{1}$School of Physics and Astronomy, University of Nottingham, 
        University Park, Nottingham, NG7 2RD, UK\\
  $^{2}$Max-Planck-Institut f\"ur extraterrestrische Physik, 
        Giessenbachstra\ss e, 85748 Garching, Germany
}

\begin{document}
  
\date{Accepted ???. Received ???; in original form ???}

\pagerange{\pageref{firstpage}--\pageref{lastpage}} \pubyear{2005}

\maketitle

\label{firstpage}

\begin{abstract}
We examine the evolution of the Tully--Fisher relation (TFR) using a
sample of $89$ field spirals for which we have
measured confident rotation velocities ($\vrot$).
This sample covers the redshift range $0.1 \la z \la 1$, with a median
of $\left<z\right>=0.33$.  The best-fitting TFR has a slope consistent with that 
measured locally, and we find no significant evidence for a change 
with redshift, although our sample is not large enough to well constrain
this.  By plotting the residuals
from the local TFR versus redshift, we find evidence that these
luminous ($M_B \la M^{\ast}_B$) spiral galaxies are increasingly
offset from the local TFR with redshift, reaching a brightening of
$-1.0\pm0.5$ mag, for a given $\vrot$, by $z \sim 1$.  This is
supported by fitting the TFR to our data in several redshift bins,
suggesting a corresponding brightening of the TFR intercept. Since
selection effects would generally increase the fraction of
intrinsically-bright galaxies at higher redshifts, we argue that the
observed evolution is probably an upper limit.

Previous studies have used an observed correlation between the TFR residuals
and $\vrot$ to argue that low mass galaxies have evolved significantly more
than those with higher mass.  However, we demonstrate that such a correlation
may exist purely due to an intrinsic coupling between the $\vrot$ scatter and
TFR residuals, acting in combination with the TFR scatter and restrictions on
the magnitude range of the data, and therefore it does not necessarily
indicate a physical difference in the evolution of galaxies with different
$\vrot$.

Finally, if we interpret the luminosity evolution derived from the TFR
as due to the evolution of the star formation rate (SFR) in these
luminous spiral galaxies, we find that SFR$(z)\propto
(1+z)^{1.7\pm1.1}$.  Notwithstanding the relatively large uncertainty,
this evolution, which is probably overestimated due to selection
effects, seems to be slower than the one derived for the overall field
galaxy population. This suggests that the rapid evolution in the SFR
density of the universe observed since $z\sim1$ is not driven by the
evolution of the SFR in individual bright spiral galaxies.
\end{abstract}

\begin{keywords}
galaxies: evolution --
galaxies: kinematics and dynamics -- galaxies: spiral
\end{keywords}

\section{Introduction}
The slope, intercept and scatter of the Tully--Fisher relation (TFR;
\citealt{TF77}) are key parameters that any successful prescription
for galaxy formation and evolution must reproduce.  As a relation
between the luminosity and rotational velocity of disc galaxies, it is
a combination of a number of fundamental galaxy properties and
relations.  These include the relation between observed rotation
velocity and total galaxy mass, that between the total and luminous
mass distributions, and the mass-to-light ratio of the galaxy's
stellar population.  The first of these is dependent upon the degree
to which the stars in the galaxy are rotationally supported and, along
with the second relation, is determined by the dynamical galaxy
formation process.  The third component, the mass-to-light ratio in a
particular photometric band, is determined by the star formation
history of the galaxy.

The TFR therefore provides a wealth of information about the
astrophysical processes involved in forming and maintaining spiral
galaxies.  In the past decade significant progress has been made
toward understanding the origins of the TFR. This involves determining
how the parameters involved conspire to maintain the relation, through
self-regulation of star-formation in the disk \citep{S97}, while
explaining the variations which lead to its intrinsic scatter.  Modern
simulations are able to reproduce the slope and scatter of the
$I$-band TFR \citep[e.g.,][]{KSW00}, respectively identifying these
with the natural range of mass and spin parameters for dark matter
haloes.  However, reproducing the TFR intercept while matching other
properties of the galaxy population is currently beyond the abilities
of semi-analytic models \citep{Cole2000}.

As well as a goal to understand in its own right, the TFR is
particularly useful as a benchmark with which to compare samples of
galaxies, in order to examine the differences between them.  Using
this method, we can gain insight into the evolution of disc galaxies
by considering the variation of the TFR with cosmic time.  In this
paper we present the $B$-band TFR of field spirals at redshifts $0.1
\la z \la 1$, and examine its evolution over the past $8$ Gyr.  We
then assess the implications of these results for the evolution of the
star formation rate (SFR) in spiral galaxies.

In Section~2 we briefly describe our observations, data reduction, and
techniques for measuring the galaxy parameters.  We present the TFR of
our field galaxy sample in Section~3, along with an examination of its
evolution with redshift, and a comparison with other studies. This is
followed by a discussion of the implications for SFR evolution in Section~4,
and finally we give our conclusions in Section~5.  Throughout we assume the
concordance cosmology, with $\Omega_{\Lambda} = 0.7$, $\Omega_{m} =
0.3$ and $H_0 = 70$ \kms~Mpc\per.  All magnitudes are in the Vega
zero-point system.

\section{Data}

The target selection, photometry and spectroscopy are described
thoroughly in \citet[hereafter \paperone]{BamfordCluster}, and
therefore only briefly summarized here.
The primary motivation for our spectroscopic observations was the comparison
of spiral galaxies in five clusters at $0.2 \la z \la 0.8$ with field
spirals at comparable redshifts.  This has been examined in \paperone.  In the
process we have observed a large number of field galaxies, by themselves
forming a useful sample for examining the evolution of the more general spiral
galaxy population.  This is the subject of the current paper.

\subsection{Target selection}

For each field, galaxies were identified and their magnitudes measured
using \sextractor{} on our $R$-band imaging.
The galaxies to be observed spectroscopically were then selected
by assigning priorities based upon the likelihood of being able to measure a
rotation curve.  Higher priorities were assigned for each of the
following: disky morphology, favourable inclination, known emission
line spectrum, and available HST data.  
A priori knowledge of the galaxies' approximate spectral type varied
depending upon the availability of such data in the literature.  
The following sources were utilized for each field, labelled by the target
cluster in that field:
MS0440: \citealt{GSLFFLH98};
AC114: \citealt{CS87,CBSES98};
A370: \citealt{DSPBCEO99,SDCEOBS97};
CL0054: \citealt{DSPBCEO99,SDCEOBS97}; P-A Duc private comm.;
MS1054: \citealt{vDFFIK00}.
This element of the selection is therefore highly heterogeneous, and certainly
not ideal.  However, while it makes an understanding of our selection function
very difficult, this does reduce the possibility of any systematic effects.

\begin{figure}
\centering
\includegraphics[height=0.45\textwidth,angle=270]%
                {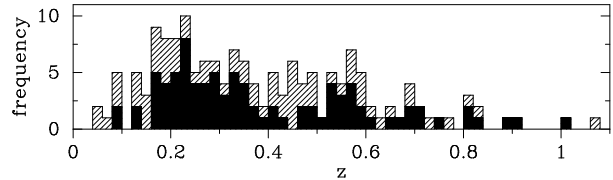}
\caption{\label{fig:zdist}
        The redshift distribution of our target galaxies.
        The hatched histogram shows the distribution of all field 
        galaxies observed with identifiable emission lines,
        and the filled area only those in our final
        TFR field sample. }
\end{figure}

For each spectroscopic mask,
slits were added in order of priority, and within each priority level
in order from brightest to faintest $R$-band magnitude.  The only
reason for a particular galaxy not being included is a geometric
constraint caused by a galaxy of higher priority level, or a brighter
galaxy in the same priority level. Often the vast majority of the mask
was filled with slits on galaxies in the highest priority levels, 
with occasional recourse to lower priority objects in order to fill
otherwise unoccupied gaps in the mask.  The redshift distribution of
our field sample is shown in Fig.~\ref{fig:zdist}.

\subsection{Photometry}
\label{sec:phot}

The photometry for this study was principally measured on HST archive
images, supplemented by our own
FORS2\footnote{\label{fn:fors}http://www.eso.org/instruments/fors}
\citep{FORS} $R$-band imaging and additional ground-based data, kindly
provided by Dr. Ian Smail.  These latter images provide additional
colour information, improving the $k$-correction by helping constrain
the galaxy SED.  The magnitudes used in this paper are measured within
SExtractor AUTO (Kron-style) apertures \citep{sextractor}, and are
corrected for Galactic extinction using the maps and conversions of
\citet*{SFD98}.

The observed apparent magnitudes were converted to apparent rest-frame
$B$-band magnitudes by a colour- and $k$-correction, determined using
the observed colour information where available and the SEDs of
\citet{AECC93}.  The absolute rest-frame $B$-band magnitudes were then
obtained assuming the concordance cosmology.  Finally the magnitudes
were corrected for internal extinction (including face-on extinction
of $0.27$ mag), following the prescription of \citet{TF85}, to give
the corrected absolute rest-frame $B$-band magnitudes, $M_B$, used in
the following analysis.  Note that both the cosmology and internal
extinction correction prescription were chosen to allow
straightforward comparison with other recent studies.

\begin{figure}
\centering
\includegraphics[height=0.40\textwidth,angle=270]%
                {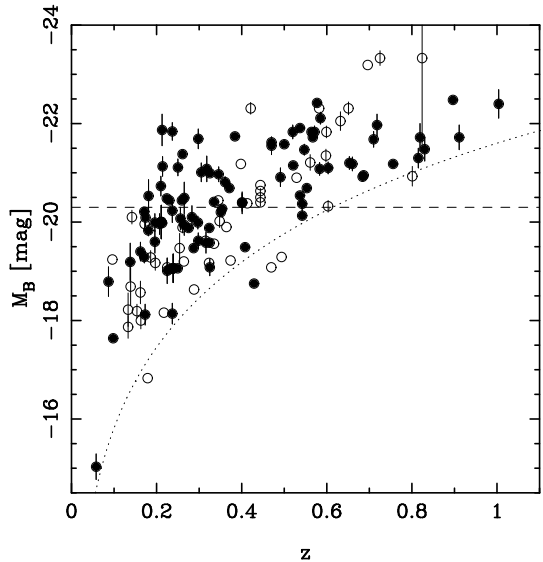}
\caption{\label{fig:m_vs_z}
        Absolute rest-frame $B$-band magnitude versus redshift
        for all our field galaxies with identifiable emission lines.
        Galaxies in our final field TFR sample are shown by filled
        points, while open points indicate those for which no emission
        lines pass our quality control criteria (see Section~\ref{sec:spec}).
        The dashed line indicates $M^{\ast}_{B}$ as calculated in the text.
        The dotted line shows the absolute magnitude corresponding to an
        apparent rest-frame $B$-band magnitude of $22.5$ mag, as a function of
        redshift in our adopted cosmology.
        }
\end{figure}

A plot of $M_B$ versus redshift is shown in Fig.~\ref{fig:m_vs_z}.
Notice how our sample is limited to brighter brighter magnitudes with
increasing redshift.  We can assess how much of the typical galaxy
population we sample by comparing with the $M^\ast$ luminosity
function parameter.  \citet{2dfLF} use data from the 2dF Galaxy
Redshift Survey to measure the $b_J$-band luminosity function, finding
$M^{\ast}_{b_J} - 5 \log (h) = -19.66\pm0.07$ mag.
We can transform this into the $B$-band using the conversion suggested
by \citet{2dfLF}, $b_J = B - 0.28(B-V)$, and the colour of a typical
galaxy in our sample (estimated from the best-fitting SEDs),
\mbox{$(B-V) = 0.52$}.  With $H_0=70$ \kms~Mpc\per, we therefore have
$M^{\ast}_{B} \simeq -20.3$ mag.  This is shown by the dashed line in
Fig.~\ref{fig:m_vs_z}.  At $z \sim 0.2$ we sample a range $\unisim
2$~mag either side of $M^{\ast}_{B}$.  By $z \sim 0.5$ this has
reduced to $(M^{\ast}_B - 2) \la M_B \la M^{\ast}_B$ mag, and at $z
\sim 0.8$ we are limited to $M_B \la (M^{\ast}_B - 1)$ mag.  We
therefore sample most of the giant spiral population below $z \sim
0.5$, but beyond this we are limited to only the brightest galaxies in
this class.  Note that this effect is less serious than in
conventional magnitude limited studies as our galaxies at high
redshifts have, on average, been observed with longer spectroscopic
integrations.  
This is because the majority of our high-redshift
galaxies are from masks targeting our more distant clusters, and thus
with longer exposure times (see Section \ref{sec:spec}).

Inclinations ($i$) and photometric scalelengths ($\rdphot$) for the disk
components of the observed galaxies were measured in bands close to $R$,
preferentially in the HST images (bands F606W, F675W or F702W), and primarily
using \gimd{} \citep{GIM2D}.  Reliable HST inclination (scalelength)
measurements were available for $47$ ($39$) per cent of the galaxies; for the
remainder inclination (scalelength) was measured on the ground-based $R$-band
imaging, again usually by \gimd{}.  These measurements are therefore separated
from the effect of the bulge component and corrected for the effect of seeing.
For a small number of galaxies the \gimd{} fit was unreliable.  In these cases
the \sextractor{} axial ratio was used, assuming an infinitely thin disk (as
does \gimd).  These inclinations were corrected by a factor determined from an
empirical comparison of \sextractor{} and \gimd{} based inclinations.  One
factor was used for all ground-based measurements, and another for the HST
measurements.

\begin{figure}
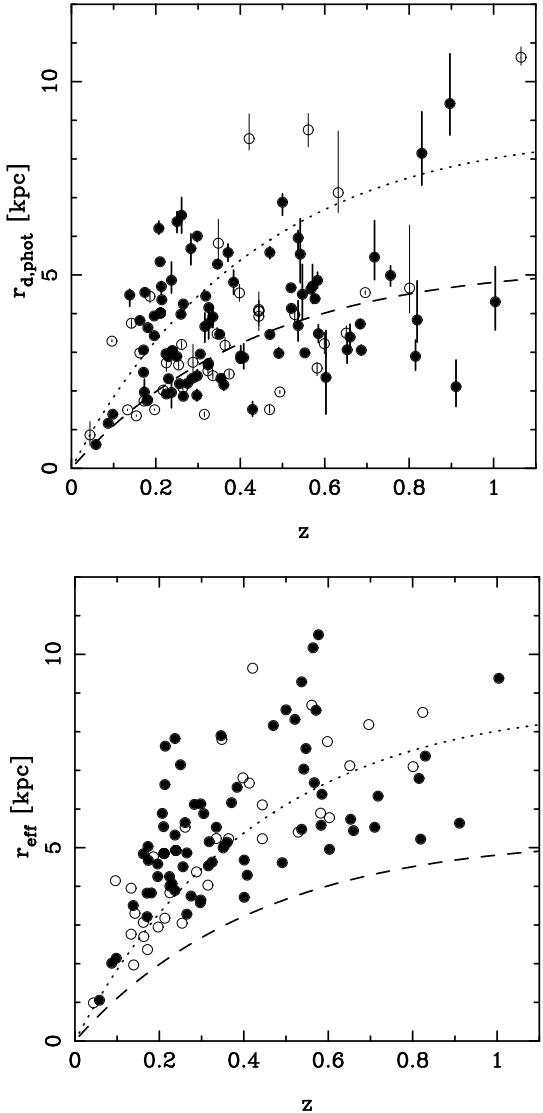

\centering
\includegraphics[height=0.40\textwidth,angle=270]%
                {fig3a.eps}
\bigskip\\
\includegraphics[height=0.40\textwidth,angle=270]%
                {fig3b.eps}
\caption{\label{fig:size_vs_z}
        Plots against redshift of (a) photometric disc scalelength, $\rdspec$ 
        and (b) effective (half-light) radius, $\reff$, both in
        kpc, for all our field galaxies with identifiable emission lines
        (excluding the supplementary MS1054 field in panel (b)).
        Galaxies in our final field TFR sample are shown by filled
        points, while open points indicate those for which no emission
        lines pass our quality control criteria (see Section~\ref{sec:spec}).
        The dotted line indicates the physical size subtended by an
        angle of $1$ arcsec, as a function of redshift in our adopted cosmology.
        The dashed line in panel (b) shows the same for an angle of $0.6$ arcsec,
        approximately equal to both the half-light radius and 
        exponential scalelength of a Gaussian with $1$ arcsec FWHM.
        }
\end{figure}

In order to further evaluate our selection function, we measure effective
radii, $\reff$, of circular apertures containing half the galaxy light.  These
were obtained from our FORS2 $R$-band imaging, the only band which is
available for nearly all of our galaxies,%
\footnote{Note that, due to full two-band \emph{HST} coverage,
FORS2 $R$-band imaging was not used by the study of MS1054 by \citet{MJetal03}.
We supplement our data set with these earlier observations, as described
later, but do not have $R$-band effective radii or surface brightness measurements
for these additional galaxies.}
using \sextractor's FLUX\_RADIUS
output.  Plots of $\rdphot$ and $\reff$ versus redshift are shown in
Fig.~\ref{fig:size_vs_z}.  Both of these plots include a dotted curve
indicating the physical size, in kpc, corresponding to an observed angle of
$1$ arcsec, as a function of redshift.  The seeing in the $R$-band imaging is
roughly $1$ arcsec FWHM, but for comparison with $\reff$ one must convert this
to a half-light radius.  A Gaussian profile with $1$ arcsec FWHM has a
half-light radius of $\unisim 0.6$ arcsec.  The measured $\reff$ are thus all
larger than the seeing, but from the correlation of the  lower bound of the
points $\reff$ distribution with redshift, the seeing is often dominating the
measurement.

In order to compare the seeing scale with $\rdphot$ one can note that if the
exponential scalelength of a point source, with a Gaussian seeing profile,
were measured, the result would be close to the radius at which a Gaussian
profile reaches $1/e$ of its peak value.  This occurs at a radius of $0.6$
times the FWHM.  However, this comparison is complicated by the inclusion of a
bulge component in the fit surface brightness model.  To some extent, though,
the fact that the $\rdphot$ distribution extends below the seeing scale, and
shows little variation in its lower limit with redshift, indicates that
$\rdphot$ is rather more independent of the seeing.

However, due to the similarity of the galaxy sizes to the seeing scale, and in
particular the impact of limited resolution on the bulge-disc decomposition,
our disc scalelengths are potentially unreliable.  This problem becomes
significantly worse for both higher redshift and intrinsically smaller
galaxies.  We therefore choose not to base any inferences on our measurements
of galaxy size, and only use them to give an indication of any biases in our
sample selection.

\begin{figure}
\centering
\includegraphics[height=0.40\textwidth,angle=270]%
                {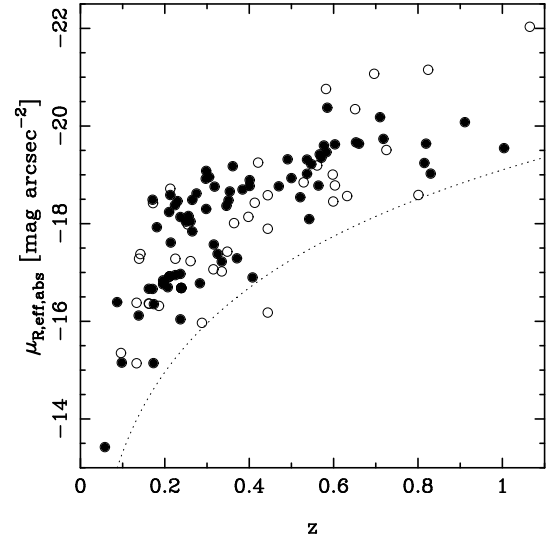}
\caption{\label{fig:sb_vs_z}
        Plot against redshift of the absolute observed-frame $R$-band surface brightness 
        within $\reff$, for all our field galaxies with identifiable emission lines
        (excluding the supplementary MS1054 field).
        Galaxies in our final field TFR sample are shown by filled
        points, while open points indicate those for which no emission
        lines pass our quality control criteria (see Section~\ref{sec:spec}).
        The dotted line indicates the absolute surface brightness corresponding
        to an observed apparent surface brightness of $25$ mag arcsec\per,
        according to our adopted cosmology.
        }
\end{figure}

Another tool to examine our sample selection is provided by the surface
brightness of our galaxies.  We calculate the apparent $R$-band average
surface brightness within $\reff$, which we denote $\mueffapp$, using
$\mueffapp = R_\rmn{eff} + 2.5 \log{2\pi\reff^2}$, where $R_\rmn{eff} = R +
2.5 \log{2}$ is the magnitude within $\reff$ by the definition of the
half-light radius, with total magnitude, $R$.  This, converted into an
absolute observed-frame $R$-band surface brightness using the distance modulus
for our adopted cosmology (denoted $\mueffabs$), is plotted in
Fig.~\ref{fig:sb_vs_z}.

A comparison of Figs.~\ref{fig:m_vs_z} and \ref{fig:sb_vs_z} reveals the two
plots to show very similar behaviour.  This, along with the apparent lack of
any redshift-dependent selection on $\rdphot$ demonstrated by
Fig.~\ref{fig:size_vs_z}(a), and the simple explanation that the correlation
in Fig.~\ref{fig:size_vs_z}(b) is due to seeing affecting the $\reff$
measurement, rather than any selection effect, implies that the surface
brightness distribution is largely a result of a magnitude-limited selection,
rather than any selection on galaxy size.

The distributions in all these plots of the galaxies which make our final TFR
sample, relative to those in the whole sample, are considered in the following
section, after a description of the final TFR sample selection criteria.

\subsection{Spectroscopy}  \label{sec:spec}

The spectroscopy for this study was observed using FORS2 \citep{FORS}, in MXU
mode, on the VLT.
In this mode, multiple slits are cut into a mask which is placed in the focal
plane.  The slits were individually tilted to align with the major axis of
each galaxy, as determined from the $R$-band pre-imaging.  Slits of 1 arcsec
width in the dispersion direction were used, with lengths that varied in order
to sample sky at the edge of each galaxy, while efficiently packing the masks
with targets.
Exposure times varied from $30$ to $210$ minutes per mask,
depending on the redshift of the cluster that formed the primary target for
each observation.
The seeing was typically $\unisim 1$ arcsec and always
below $1.2$ arcsec.
These data and their reduction are described more completely in \paperone,
and are therefore only briefly summarised here.

The data were reduced in the usual manner (\citealt{MJthesis}; Bamford
thesis in preparation) to produce straightened, flat-fielded,
wavelength-calibrated and sky-subtracted 2d-spectra for each galaxy.
From these, individual sky- and continuum-subtracted emission line
`postage stamps' were extracted.
The main emission lines observed were [OII]$\lambda 3727$, H$\beta$ and
[OIII]$\lambda\lambda4959,5007$, with H$\alpha$ but no [OII]$\lambda 3727$ for
nearby galaxies ($z \la 0.3$, depending upon the position of the slit in the
mask).  In order to measure the rotation velocity ($\vrot$) and emission scale
lengths ($\rdspec$) we fit each emission line independently using a synthetic
rotation curve method based on \elfitd{} by \citet{SP98,SP99}, and dubbed
\elfitpy.  In this technique a model emission line is created for particular
sets of parameters, assuming a form for the intrinsic rotation curve, an
exponential surface-brightness profile, and given the galaxy inclination,
seeing and instrumental profile.  The intrinsic rotation curve assumed here is
the `universal rotation curve' (URC) of \citet{PS91}, with a slope weakly
parametrized by the absolute $B$-band magnitude, $M_B$.  
From a comparison of the results from fits using both flat and `universal'
intrinsic rotation curves, adopting a flat rotation curve leads to values of
$\vrot$ $\unisim 10$ \kms{} lower than found with the URC.  The choice of
rotation curve does not appear to affect the conclusions of this study.  A
Metropolis algorithm is used to search the parameter space to find those which
best fit the data, and to determine confidence intervals on these parameters.
Images of model lines with the best-fitting parameters are also produced for
comparison with the data.

In order to produce a single value of $\vrot$ and $\rdspec$ for each
galaxy, the values for the individual lines are combined by a weighted
mean, as described in \paperone.

A significant fraction of the emission lines identified display
dominant nuclear emission, or asymmetries in intensity, spatial extent
or kinematics.  In severe cases these departures from the assumed
surface brightness profile and intrinsic rotation curve mean that the
best-fitting model is not a true good fit to the data.  A similar
situation can occur for very low signal-to-noise (\snr) lines, where
an artifact of the noise overly influences the fit. More concerning is
the case of very compact lines, where the number of pixels is on the
order of the number of degrees of freedom in the model, and hence an
apparently good fit is obtained despite a potentially substantial
departure from the assumed surface brightness profile.

In order to eliminate such `bad' fits a number of quality tests are
imposed, based on a measure of the median \snr{} and a robust $\chi^2$
goodness-of-fit estimate.  In addition, the emission lines were
`traced' by fitting a Gaussian to each spatial column of pixels, and
the region determined for which the trace is reliable.  The distance
from the continuum centre to where the line could no longer be
reliably detected above the noise we term the \emph{extent}.
Additionally, quantities describing the asymmetry, in terms of extent
and kinematics, and the flatness of the line at maximum extent were
formulated.  Unfortunately, a satisfactory set of quantitative
criteria, to exclude `bad' fits while efficiently retaining `good'
fits, could not be found.  Therefore, we supplemented cuts on \snr and
$\chi^2$ with a visual inspection of every emission line, supported by
the various quality tests just described.  In the process galaxies
with apparently dominant nuclear emission were excluded, as were lines
producing zero $\vrot$ estimates and those clearly inconsistent with a
number of other lines for the same galaxy.

As in \paperone, we have supplemented our data with that from
\citet{MJetal03}.  This is also FORS2 MXU data, reduced in a very
similar way to that described above.  In order to consistently combine
the MS1054 data with this sample, the emission line postage stamps for
the MS1054 galaxies have been re-fit using \elfitpy{} and the same
line quality criteria applied.  

After all these quality checks, our full field galaxy TFR sample
contains $89$ galaxies, with a mean of $\unisim 2.3$ emission lines
contributing to the measurements for each galaxy.
This sample covers the redshift range $0.1 \la z \la 1$, with a median
redshift of $\left<z\right>=0.33$, as show in Fig.~\ref{fig:zdist}.

Figures \ref{fig:m_vs_z}, \ref{fig:size_vs_z} and \ref{fig:sb_vs_z} show the
distributions of $M_B$, $\rdphot$, $\reff$ and $\mueffabs$ versus redshift for
all field emission line galaxies observed.  Those in our final TFR sample are
marked by filled points, while those which are not, i.e., for which none of the
emission line fits passed our quality criteria, are shown by open points.
There is no noticeable difference between the distributions of selected and
rejected galaxies in these plots.  This indicates that our quality selection
criteria are not biasing our sample from the point of view of the galaxies'
broadband photometric properties, beyond those baises inherent to our initial
spectroscopy observational selection procedure.

\section{The Tully--Fisher relation}

\subsection{Basic fit} \label{sec:TFR_fit}

\begin{figure}
\centering
\includegraphics[height=0.45\textwidth,angle=270]%
                {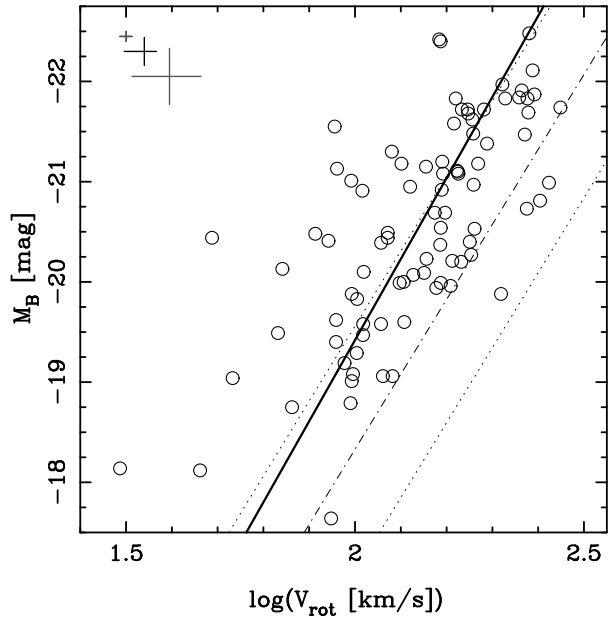}
\caption{\label{fig:tf}
        The Tully--Fisher relation for our full field TFR sample.
        The fiducial local relation of \citet{PT92} is marked by the dot-dashed
        line, with its $3\sigma$ intrinsic scatter delimited by dotted lines.
        A weighted least-squares fit to all the points is shown by the
        solid line, see the text for details.
        The error bars in the top left corner indicate the 10th-, 50th-
        and 90th-percentiles of the (broadly unimodal) distribution of
        uncertainties on the displayed points.
        }
\end{figure}

Our TFR for distant field galaxies is given in Fig.~\ref{fig:tf}.  The
solid line is an `inverse' fit to the data,
\begin{equation}
\log{\vrot} = a + bM_{B}
\end{equation}
found by minimising the weighted squared deviations in $\log{\vrot}$.
The weights applied to each point are
\begin{equation}
w_i = \slantfrac{1}{\sigma_i^2} \;,
\end{equation}
where
\begin{equation}
\sigma_i^2 = \sigma_{\log{\vrot},i}^2 + b^2 \sigma_{M_{B},i}^2 + \sigma_\rmn{int}^2 \;.
\end{equation}
Here $\sigma_{\log{\vrot},i}$ and $\sigma_{M_{B},i}$ are the errors on
each point derived from the reduction process, and
$\sigma_\rmn{int}$ is the intrinsic scatter of the TFR, which is
allowed to vary such that the reduced chi-squared statistic,
$\chi_\rmn{r}^2$, is unity.  This was achieved by iteration, each
time recalculating the weights using the new value of $b$, and
$\sigma_\rmn{int}$ determined via the recurrence relation
\begin{equation} \label{eqn:sigma_int}
\sigma_{\rmn{int},{j+1}}^2 = \sigma_{\rmn{int},{j}}^2\, \chi_\rmn{r}^{2\alpha}
        \; ; \quad \rmn{with}\ \alpha > 0 \;,
\end{equation}
where, as usual for two degrees of freedom, the reduced-chi-squared statistic is
\begin{equation}
\chi_\rmn{r}^2 = \frac{\chi^2}{n-2}\;,\quad 
\chi^2 = \sum_i{\left(\frac{\log{\vrot}_i - a - bM_{B,i}}{\sigma_i}\right)^2} \;.
\end{equation}
The (weighted) total scatter, $\sigma_\rmn{total}$, is calculated by
\begin{equation}
\sigma_\rmn{total}^2 = \sum_i{w_i\left(\log{\vrot}_i - a - bM_{B,i}\right)^2} \left/ 
                          \sum_i{w_i} \right. \;.
\end{equation}

Notice that when $\chi_\rmn{r}^2$ is at the desired value of unity,
the $\sigma_\rmn{int}^2$ recurrence relation
(\eqnref{eqn:sigma_int}) reduces to $\sigma_{\rmn{int},{j+1}}^2 =
\sigma_{\rmn{int},{j}}^2$, as we would want.  When $\chi_\rmn{r}^2 >
1$ this implies $\sigma_\rmn{int}^2$ is too small, and thus it is
increased for the next iteration.  Alternatively, when $\chi_\rmn{r}^2
< 1$ this implies $\sigma_\rmn{int}^2$ is too large, and it is
therefore decreased for the next iteration.
The value of $\alpha$ may be tuned to minimise the number of iterations
before convergence is achieved;  $\alpha = \frac{2}{3}$ was found
to work well.

The fit converges in $\unisim 4$ iterations to give the TFR (converted
back into the `forward' form):
\begin{equation} \label{eqn:tf}
M_{B} = (-8.1\pm0.8) \cdot \log{\vrot} \:+\: (-3.2\pm2.0)
\end{equation}
with $\sigma_\rmn{total} = 1.0$ mag and $\sigma_\rmn{int} = 0.9$ mag.

The `inverse' fit is often used to avoid a potential bias because of
the apparent magnitude limited nature of a TFR sample.  At faint
apparent magnitudes, points are preferentially selected above the real
TFR.  For samples over a narrow range of distance moduli, e.g., in
local cluster studies, the nature of this bias is to cause an apparent
flattening of the TFR slope.  While the dependence of $\vrot$ on $M_B$
still causes an effect, \citet{Willick94} has shown that using an
`inverse' fit reduces the bias by a factor of five compared with the
conventional `forward' fit.  As this study covers a wide range of
distance moduli, apparently faint galaxies are found over the whole
range of absolute magnitude, depending on their redshift.  Any bias on
the best-fitting TFR should therefore be weakened, and take the form
of an intercept shift rather than a flattening of the slope.  This is
particularly true given the more random priority-based, rather than
magnitude-based, selection. However, to minimise any such effect, and
allow us to confidently use the same fitting method to compare
differently selected samples, we choose to work with the `inverse'
fit.

We weight our fit to make full use of the data, and avoid the
influence of unreliable points, while the inclusion of an intrinsic
scatter term prevents points with small errors from dominating the
fit, and allows us to estimate this useful parameter.

The thin lines in Fig.~\ref{fig:tf} indicate a fiducial local field
TFR.  This is derived from the TFR of \citet[hereafter PT92]{PT92},
with a zero-point adjustment because PT92, while otherwise using the
internal extinction correction of \citet{TF85}, do not include the
$0.27$ mag of face-on extinction that is applied to our data.  The
fiducial PT92 TFR, adapted to our internal extinction correction, is
thus:
\begin{equation} \label{eqn:PT92}
M_{B}^\rmn{PT92}(\vrot) = -7.48\log{\vrot} - 3.37 \; .
\end{equation}

PT92 obtained this relation by an `inverse' least-squares fit,
minimising the residuals in rotation velocity.  This is the same
method we have used to fit our TFR, except that our fit is weighted
and includes the intrinsic scatter in these weights.  The weighting
should not cause any bias, and hence our fitting method is comparable
to that used to produce the fiducial TFR of PT92.

There is clearly a significant offset between the TFRs of PT92 and
this study. This may not be entirely an evolutionary effect, and some
part of it is likely due to the different manner in which we measure
the rotation velocities and magnitudes compared with PT92.  Note that
in generating the fiducial relation we have assumed that $\vrot = 0.5
W^i_R$, where $W^i_R$ is the fully-corrected HI velocity width
measured by PT92.

A further issue is whether or not the absolute calibration of the PT92
TFR is correct.  This is based upon Cepheid and RR Lyrae distances to
six local calibrator galaxies, and hence dependent upon the
calibration of the whole extra-galactic distance scale at the time.
There is also the issue of how representative this small number of
local calibrator galaxies are compared with the whole PT92 TFR sample.
\citet[hereafter P94]{Pierce94} uses the PT92 TFR to calibrate
supernovae distances and derive $H_0 = 86 \pm 7$ \kms~Mpc\per.  If, as
is suggested, the primary cause of uncertainty in this value is the
absolute TFR calibration, then we could potentially use the modern,
significantly more accurate value of $H_0$ to correct this.  We could
therefore derive a more accurate TFR absolute calibration by using our
cosmologically measured $H_0$, a reversal of the traditional method.

The current best estimate of $H_0$ is provided by a combination of
data from WMAP and a number of other surveys: $H_0 = 71$\asymerr{3}{4}
\kms~Mpc\per{} \citep{WMAPcosmo}.  The difference in distance modulus
between this $H_0$ and that derived by P94 is $0.5\pm0.1$ mag, in the
sense of moving the TFR to brighter magnitudes.  The errors given only
include those on the WMAP $H_0$ measurement, as the P94 errors are
dominated by the uncertainty on the distance scale, which we are
replacing.  It can therefore be argued that a correction of $-0.5$ mag
should be applied to the intercept of the PT92 TFR.

Fortunately in this study the uncertainty on the intercept of the
fiducial local TFR is of little concern.  Our sample covers a wide
range in redshift, selected and analysed in a homogeneous manner.  We
can therefore examine the TFR evolution using only this study's data,
without recourse to external work.

In future studies it may be wise to consider using the more recent
$B$-band TFR of \citet{TP00} as a comparison. This relation has an
absolute calibration based on Cepheid distances to $24$ galaxies and
implies a value of $H_0 = 77 \pm 8$, more consistent with the WMAP
result.
The work of \citet{Verheijen01} provides another
more recent local comparison TFR, which has already been used by some groups.
However, these studies use the internal extinction correction scheme of
\citet{TPHSVW98}, which has a strong dependency upon galaxy luminosity
(or alternatively $\vrot$).  While this form of the internal
extinction should be more accurate, it will require care as the
dependency on luminosity (or $\vrot$) is calibrated locally, and may
not be valid for distant galaxies.

The internal scatter we measure, $\sigma_\rmn{int} = 0.90$ mag, is
considerably larger than the $\sim 0.4$ mag generally found for local
samples \citep[e.g][]{PT92,DGHCH99}.  However, most local studies are
focused upon using the TFR as a distance indicator.  They therefore
impose very strict selection criteria, e.g., requiring undisturbed
late-type spirals, in an effort to produce as strong a correlation as
possible.  At high redshift we do not have the luxury of abundant
data, and so cannot impose such strict criteria.  For example, it is
harder to identify galaxies with slightly disturbed rotation curves
and morphologies, making rejection of such objects impossible.  In
addition, corrections for known, locally-calibrated correlations,
e.g., between TFR intercept and morphological type, are often applied
to local samples.  However, such corrections are not yet calibrated at
high-redshift and so cannot be applied to our data.  The variations
caused by minor disturbances, differing sample selection, and varying
corrections may be responsible for the entire increase in intrinsic
scatter.  However, physical effects, such as an increased
stochasticity of star-formation, may also contribute.

\subsection{TFR evolution with redshift}  \label{sec:evolution}

The slope of the TFR we measure for the whole field galaxy sample is
very similar to that found locally by PT92, and consistent within our
errors.  Note that we have used a comparable fitting method to PT92,
so this statement is valid.  The precise slope measured is very much
dependent on the fitting method employed, and may explain why some
studies find apparently different slopes.  Our fit suggests there is
little change in the TFR slope with redshift.  We can attempt to
evaluate this further by fitting redshift sub-samples of our data.

As noted above, any difference in intercept between our sample and the
local relation of PT92 may not be a real effect.  However, by fitting
redshift sub-samples we may investigate any evolution of this offset
purely within our own sample.

\begin{figure}
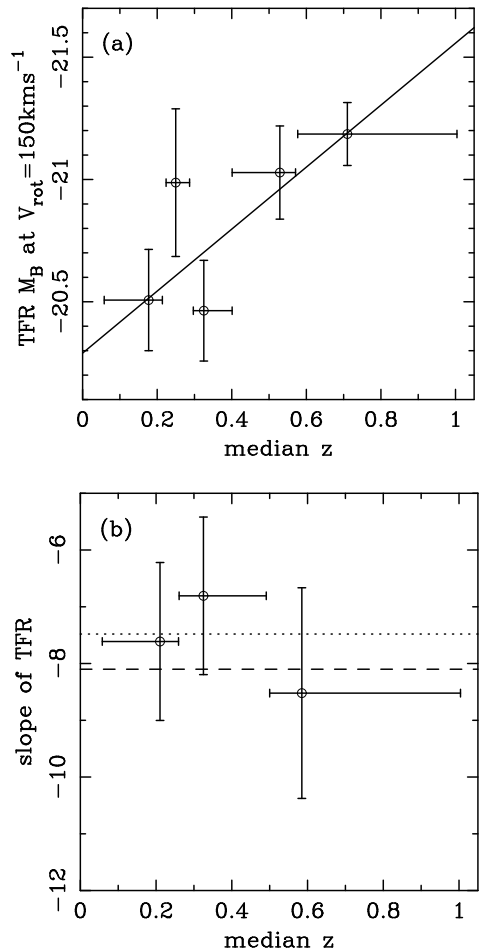

\centering
\includegraphics[height=0.35\textwidth,angle=270]%
                {fig6a.eps}
\bigskip\\
\includegraphics[height=0.35\textwidth,angle=270]%
                {fig6b.eps}
\caption{\label{fig:tf_evolution}
        Plots showing the evolution of the slope and intercept of 
        the best-fitting TFR in several redshift bins, indicated by the
        horizontal error bars.
        Panel (a) gives the intercept (at the median $\vrot$ of the
        full sample) for TFR fits to five redshift subsamples of the data.
        Each redshift bin contains $\unisim 18$ points. The line shows
        a weighted least-squares fit to the points.
        For these fits the slope was held fixed to the PT92 value. 
        Panel (b) plots the slope for TFR fits to three redshift subsamples
        of the data. Each redshift bin contains $\unisim 30$ points.
        The horizontal lines show the local TFR slope of PT92 (dotted)
        and the slope from \eqnref{eqn:tf}, the fit to our whole field 
        TFR sample (dashed).
        }
\end{figure}

In order to examine the intercept evolution we divide the sample in to
five bins, each of $18$ points ($17$ for the highest redshift bin).
The TFR is fit for each bin using the usual method, but with the slope
constrained to the PT92 value (and hence very similar to the value for
our whole sample) to isolate the changes in intercept.  In panel (a)
of Fig.~\ref{fig:tf_evolution} we plot these results, as the $M_B$
value of the best-fitting TFR at the median $\vrot$ of the full
sample.  Also plotted is a conventional weighted least-squares fit to
the points.  There is reasonably clear evidence for a brightening of
the TFR intercept with redshift of $\unisim 1$ mag by $z = 1$,
corresponding to a factor of $\unisim 2.5$ in luminosity at a fixed
rotation velocity.

At high redshift, however, we only sample the bright end of the galaxy
luminosity function, and hence preferentially select objects which
have been brightened.  In particular we require a certain emission
line flux to measure $\vrot$, and hence the limiting emission line
luminosity of our sample will increase with redshift.  As emission
line luminosity increases with SFR, we will therefore sample galaxies
with higher SFR at higher redshifts, which in turn implies brighter
$B$-band magnitudes for the high-redshift galaxies.  The evolution of
the TFR intercept that we measure is therefore probably an upper limit
on the true brightening, e.g., at a given galaxy mass.

To investigate any evolution in the TFR slope we divide the full
sample into only three redshift bins, as the slope requires more
points to constrain it.  Each bin thus contains 30 points (29 for the
highest redshift bin).  A TFR was fit to each sub-sample in the same
way as for the full sample, and the results are shown in panel (b) of
Fig.~\ref{fig:tf_evolution}.  Due to the small numbers of points in
each bin, the errors on the slope are quite large.  Given these
errors, and the restricted $M_B$ range of the data in the highest
redshift bin, no strong constraints can be inferred.  However, the
slope in each bin is consistent with no evolution of this parameter
with redshift.

\begin{figure}
\centering
\includegraphics[height=0.45\textwidth,angle=270]%
                {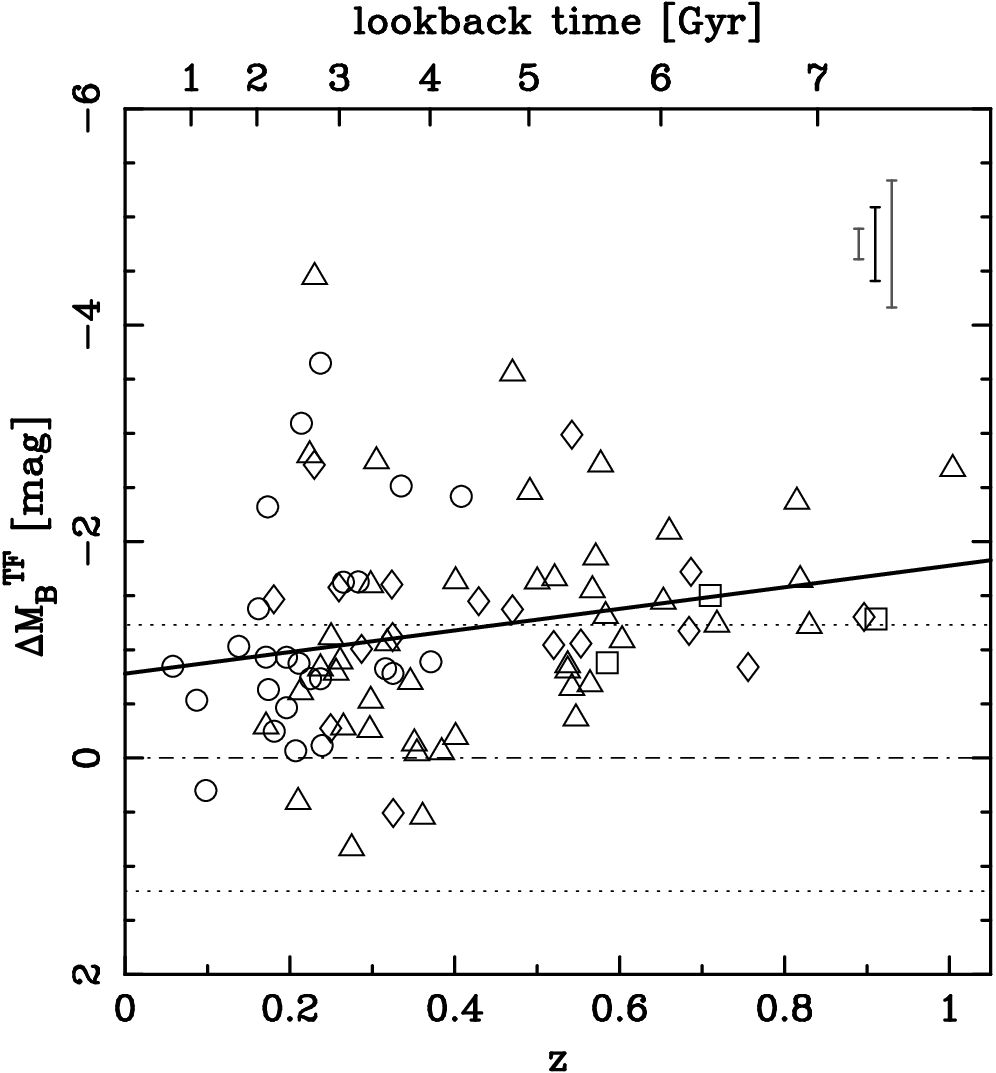}
\caption{\label{fig:tf_res_z_wrt_pt92}
        The residuals, $\dtf$, from the fiducial PT92 TFR plotted against redshift
        and look-back time. 
        Galaxies with $\mueffabs < -20.0$ are denoted by squares,
        those with $-20.0 < \mueffabs < -18.0$ are shown by triangles,
        circles denote those with $\mueffabs > -18.0$, and galaxies
        with no $\mueffabs$ measurement (i.e., those from the supplementary
        MS1054 data) are indicated by diamonds.
        The fiducial local relation of PT92 is again marked by the 
        thin dot-dashed and dotted lines.
        A weighted least-squares fit to all the points is shown by the
        solid line, see the text for details.
        The error bars in the top left corner indicate the 10th-, 50th-
        and 90th-percentiles of the (broadly unimodal) distribution of
        uncertainties on the displayed points.
        }
\end{figure}

In order to further investigate changes in the intercept of the TFR
via the offsets of individual galaxies, it is helpful to work with the
residuals from the fiducial TFR:
\begin{eqnarray} \label{eqn:dtf}
\dtf &=& M_{B} - M_{B}^\rmn{PT92}(\vrot) \nonumber \\
     &=& M_{B} - \left( -7.48\log{\vrot} - 3.37 \right) \; .
\end{eqnarray}
We can evaluate an evolution of the TFR with redshift by looking at a
plot of these residuals versus redshift, as shown in
Fig.~\ref{fig:tf_res_z_wrt_pt92}.  A trend in $\dtf$ with redshift,
such that more distant galaxies are brighter for a given rotation
velocity, is apparent.  Fitting the data in a similar manner to that
described for the TFR in Section~\ref{sec:TFR_fit} (minimising residuals in
$\dtf$) produces the relation
\begin{equation} \label{eqn:dtf_z}
        \dtf = (-1.0\pm0.5) \cdot z + (0.8\pm0.2) \ \rmn{mag}
\end{equation}
with $\sigma_\rmn{total} = 0.9$ mag and $\sigma_\rmn{int} = 0.8$ mag.
This is in good agreement with the more qualitative result from
Fig.~\ref{fig:tf_evolution}.  Once again, however, this is most likely
an upper limit on the true luminosity evolution experienced by giant
spirals in the field, due to our preferential selection of the
brightest objects at high redshift.

In Fig.~\ref{fig:tf_res_z_wrt_pt92} there is clearly some asymmetric
scatter to brighter offsets, which decreases with redshift.  This may
be due, at least in part, to the fact that there is a larger scatter
in the TFR around $\vrot \sim 100$ \kms.  Only brighter galaxies, and
hence those with $\vrot \ga 150$ \kms{} and thus lower TFR scatter,
are observed at high redshifts.  However, this does not seem able to
account for the whole effect, and there is a hint that this reduction
in scatter with redshift is real.

The points in Fig.~\ref{fig:tf_res_z_wrt_pt92} are shown using different
symbols depending upon absolute surface brightness, $\mueffabs$.  There
appears to be no obvious difference in the $\dtf$ distribution of high and low
surface brightness galaxies at a given redshift.  This argues against the
variation in scatter being simply a consequence of sampling different surface
brightness ranges at different redshifts.  It also implies that the varying
surface brightness selection function has relatively small impact on the
evolution we measure.

As mentioned in the previous section, there is a systematic offset
between our TFR and that of PT92, even when the evolution with
redshift is taken into account (as indicated by the zero-point of
\eqnref{eqn:dtf_z} being inconsistent with zero).  The earlier
discussion of the uncertainties in the intercept of the PT92 TFR also
applies here.  Note that the `cosmological' correction of $-0.5$ mag
suggested in Section \ref{sec:TFR_fit} would bring the zero-points
into much closer agreement.  In addition, there are a variety of other
possible explanations for the offset.  For example, the asymmetric
scatter to brighter $M_B$ for a given $\vrot$, seen particularly at
lower redshifts, may influence the fit away from the PT92 relation.
This scatter may be due to galaxies that would not have been included
in the PT92 sample, for example due to interactions, or that are less
prevalent locally.  A more rapid evolution in the past $\unisim 2$
Gyr, which is not well constrained by our data, could also be
responsible.

\subsection{TFR residuals versus $\bmath \vrot$}
\label{sec:dtf_vrot}

\begin{figure*}
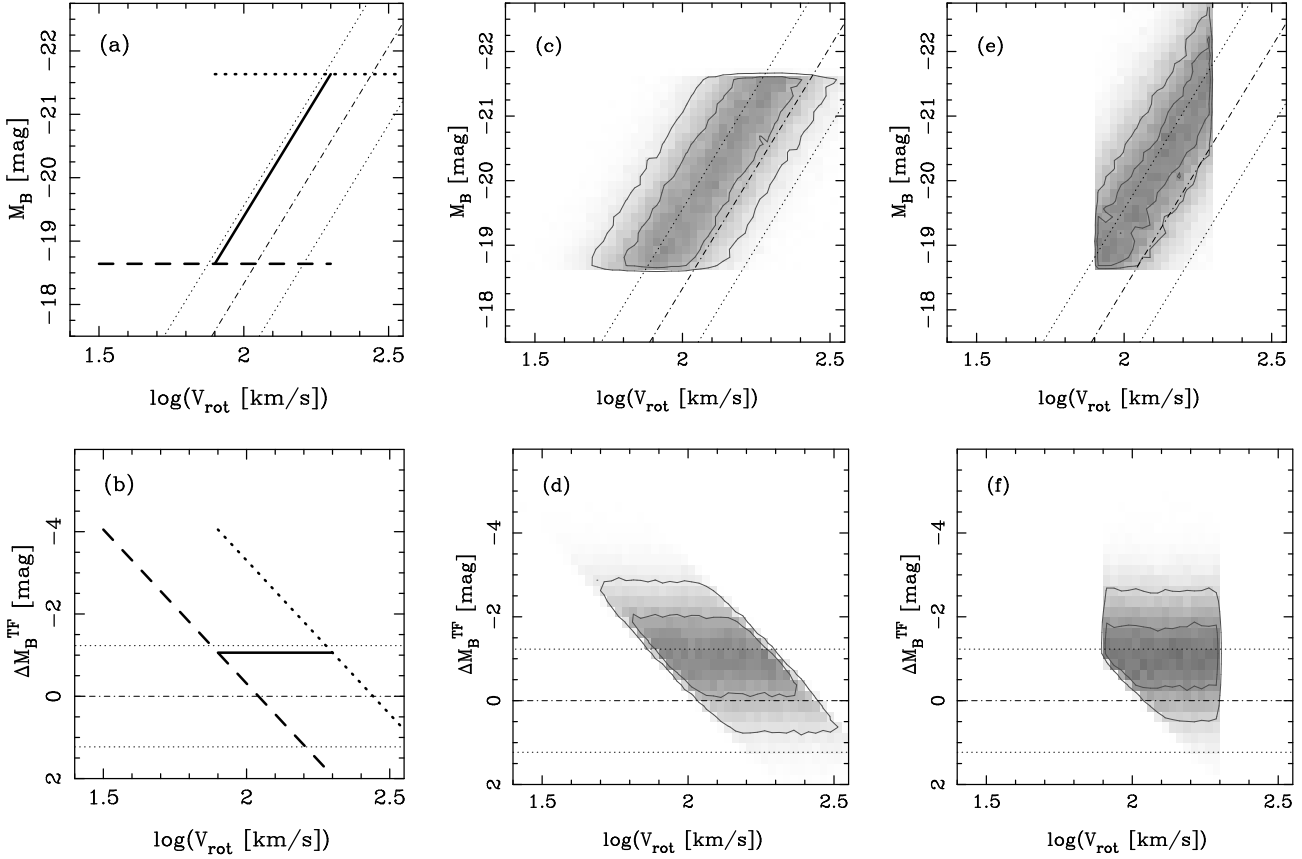

\centering
\hspace*{\stretch{1}}
\includegraphics[height=0.3\textwidth,angle=270]%
                {fig8a.eps}
\hspace*{\stretch{2}}
\includegraphics[height=0.3\textwidth,angle=270]%
                {fig8c.eps}
\hspace*{\stretch{2}}
\includegraphics[height=0.3\textwidth,angle=270]%
                {fig8e.eps}
\hspace*{\stretch{1}}
\bigskip\\
\hspace*{\stretch{1}}
\includegraphics[height=0.3\textwidth,angle=270]%
                {fig8b.eps}
\hspace*{\stretch{2}}
\includegraphics[height=0.3\textwidth,angle=270]%
                {fig8d.eps}
\hspace*{\stretch{2}}
\includegraphics[height=0.3\textwidth,angle=270]%
                {fig8f.eps}
\hspace*{\stretch{1}}
\caption{\label{fig:tf_res_vrot_example}%
  A simple demonstration of the intrinsic correlation between the TFR
  residuals from the fiducial PT92 TFR, ie., $\dtf$, and $\vrot$.  The fiducial
  local relation of PT92 is again marked by the thin dot-dashed and dotted
  lines.  Panels (a) and (b) show instructive diagrams of the TFR and $\dtf$
  versus $\log{\vrot}$ plot, respectively.  Each line in panel (b)
  corresponds to the same style of line in panel (a).
  Panels (c) and (d) show the distribution of points, as greyscale and
  contours, for simulated galaxies with true $\log(\vrot)$ uniformly
  distributed on the range $1.9$--$2.3$, true $M_B$ assigned to each galaxy
  corresponding to the PT92 TFR, with a constant offset to make it more easily
  comparable to our intermediate redshift data.  Gaussian scatter with 0.133
  dex standard deviation (corresponding to 1.0 mag in terms of $M_B$ assuming
  the PT92 TFR slope) has been added to generate the observed $\log{\vrot}$.
  No scatter is added to the $M_B$ values.
  Panels (e) and (f) show the distribution of simulated galaxies with true
  properties created in the same manner as the previous panels.  However, in
  generating the observed values in this case, no scatter has been added to
  $\log{\vrot}$, while Gaussian scatter with 1.0 mag standard deviation is
  added to the $M_B$.  In addition, to demonstrate the effect of a magnitude
  cut, points with observed $M_B > -18.64$ have been removed from the sample.
  See the text for a more detailed discussion. }
\end{figure*}

\begin{figure*}
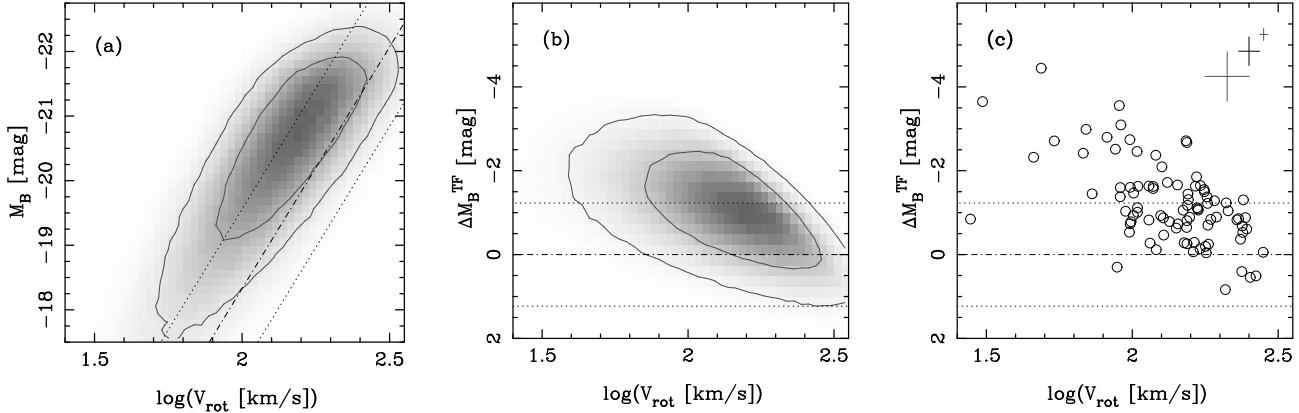

\centering
\hspace*{\stretch{1}}
\includegraphics[height=0.3\textwidth,angle=270]%
                {fig9a.eps}
\hspace*{\stretch{2}}
\includegraphics[height=0.3\textwidth,angle=270]%
                {fig9b.eps}
\hspace*{\stretch{2}}
\includegraphics[height=0.3\textwidth,angle=270]%
                {fig9c.eps}
\hspace*{\stretch{1}}
\caption{\label{fig:tf_res_vrot_sim}
  A more detailed demonstration of the intrinsic correlation between $\dtf$ and
  $\vrot$.  The fiducial local relation of PT92 is again marked by the thin
  dot-dashed and dotted lines.
  Panels (a) and (b) show the distribution of points in the TFR and $\dtf$
  versus $\log{\vrot}$ plot, respectively, for a realistic simulated sample
  of galaxies, as described in the text.
  This may be compared with panel (c), a plot of $\dtf$ versus $\log{\vrot}$
  for our real field TFR sample.  Note the clear correlation, which, as shown
  by the simulations, may be explained purely by the intrinsic correlation
  between $\dtf$ and $\log{\vrot}$, along with the effect of scatter in the
  TFR combined with a restriction on the magnitude range of the data.
  This correlation, therefore, does not necessarily imply that the data
  prefers a slope of the TFR different to that of the chosen fiducial TFR, and
  is thus not adequate evidence for such a real, physical effect. }
\end{figure*}

It may be thought that plotting the TFR residuals, $\dtf$, versus $\vrot$
would provide a test for a change of slope.  However, this must be treated
very carefully, as the two variables are intrinsically correlated through the
scatter in $\vrot$.  \citet[hereafter B04]{Bohmetal04} use this plot to argue
that galaxies with lower $\vrot$, and hence lower mass, are offset further
from the local TFR than more massive galaxies.  However, a correlation between
$\dtf$ and $\vrot$ does not necessarily imply such a result.  The fact that
the two variables are not independent (cf. \eqnref{eqn:dtf}), combined with
scatter in the TFR and restrictions on the magnitude range of the data, causes
an intrinsic correlation, even in the absence of any true difference in TFR
slope.

In order to show this in more detail, Fig.~\ref{fig:tf_res_vrot_example} gives
a pedagogical illustration of the effect of scatter and magnitude cuts on the
distribution of points in the TFR and \mbox{$\dtf\!\!$~--~$\log{\vrot}$} plots.
Panels (a) and (b) show three corresponding lines in the TFR and \mbox{$\dtf\!\!$~--~$\log{\vrot}$} plot, respectively.  The thick solid line is an example
ideal TFR, corresponding to the PT92 relation with a constant magnitude offset
of $-1.06$ mag, the median $\dtf$ of our data, in order to allow more direct
comparison of the examples with our data.  This example TFR is limited to the
range $1.9 < \log{\vrot} < 2.3$, and hence $-18.64 > M_B > -21.63$.  The thick
dashed and dotted lines indicate loci of constant $M_B$ at the faint and
bright limits of the example TFR, respectively.  In panel (b), the plot of
$\dtf$ versus $\log{\vrot}$, the example TFR becomes a (relatively short)
horizontal line, while the loci of constant $M_B$ are lines with a slope
negative that of the example TFR, as is obvious from consideration of
\eqnref{eqn:dtf}.

To show the transformation of distributions of points in the TFR plot to the
\mbox{$\dtf\!\!$~--~$\log{\vrot}$} plane we construct samples of galaxies with
simulated properties.  These are designed to cover parameter ranges fairly
similar to our data, but in order to keep our arguments straightforward, we
initially keep these simulations highly simplified.  More realistic
simulations are considered later.

Each simulated galaxy is assigned a `true' rotation velocity from a random
uniform distribution in $\log{\vrot}$ between $1.9$ and $2.3$, and a `true'
magnitude corresponding to this `true' rotation velocity using the example TFR
described above.  The `true' galaxy properties are thus uniformly randomly
distributed on the solid line in panel (a) of
Fig.~\ref{fig:tf_res_vrot_example}.  The distribution of `true' properties is
the same irrespective of whether the initial property assigned is
$\log{\vrot}$ or $M_B$.  Scatter, in terms of measurement errors or intrinsic
scatter in the TFR, may then be added to either the `true' rotation
velocities, magnitudes, or both, to produce simulated `observed' values of the
parameters.

The effects of measurement error and intrinsic scatter on the distribution of
points are equivalent for these examples.  Differences between the two can only
arise as artifacts of the simulation method.  Measurement error must always be
applied after both parameters have been assigned, while intrinsic scatter may
be applied between the production of the first parameter, e.g., $\log{\vrot}$
and the assignment of the corresponding second parameter, e.g., $M_B$.  If the
intrinsic scatter is only applied to the value of the first-assigned parameter
that is then used to generate the second parameter, then this is equivalent to
a measurement scatter applied directly to the second parameter.  If the
intrinsic scatter is only applied to the first parameter after the second
parameter has been generated, then this is equivalent to measurement error on
the first parameter.  Alternatively, if the intrinsic scatter is applied to
the `true' value of the first-assigned parameter, which is then used to
generate the second parameter and also used as a basis for the `observed'
value of the first parameter, then this is effectively an alteration of the
initial distribution from which the first parameter is derived, as well as the
addition of a scatter in the second parameter which is indistinguishable from
a measurement error.  As we wish to be able to control the distribution of the
`true' parameters independently of the inclusion of scatter, we choose to
implement intrinsic scatter as offsets applied following the generation of
both `true' parameters, in which case it is identical to a scatter due to
measurement error.  For the following simple examples we therefore do not
distinguish between the two.  All random scatters applied in these
simulations are based on Gaussian distributions with the specified standard
deviation.

In panel (c) of Fig.~\ref{fig:tf_res_vrot_example} we plot the distribution of
$10^5$ simulated points, as greyscale and contours, generated by the procedure
described above, and with the inclusion of 0.133 dex of scatter in
$\log{\vrot}$ (corresponding to 1.0 mag in terms of $M_B$ assuming the PT92
TFR slope).  As no scatter has been added to the magnitude coordinates, the
distribution is restricted to a sharply defined range of $M_B$.  Panel (d)
shows the corresponding distribution in the \mbox{$\dtf\!\!$~--~$\log{\vrot}$} plane.
The extension of the points parallel to the fiducial TFR produces a horizontal
extension in this plot.  The scatter in $\log{\vrot}$ spreads the distribution
in both the horizontal and vertical direction, along lines with slope equal to
minus the TFR slope.  The restrictions in $M_B$, due to the initial choice of
the `true' TFR points' distribution, translate into sharply defined sloping
cut-offs in the \mbox{$\dtf\!\!$~--~$\log{\vrot}$} plot, corresponding to the dashed
and dotted lines shown in panels (a) and (b) of the same figure.

Panel (e) of Fig.~\ref{fig:tf_res_vrot_example} shows, in the same manner as
panel (c), the distribution of another $10^5$ simulated points.  These points
were generated by the same procedure, but this time with 1.0 mag of scatter in
$M_B$, and none in $\log{\vrot}$.  In addition, in this panel we demonstrate
the effect of a magnitude cut, by rejecting points with observed $M_B >
-18.64$ (corresponding to the faint end of the simulated `true' $M_B$
distribution) from the sample.  Without a magnitude cut, the edges of the
distribution are defined by sharp vertical lines of constant $\log{\vrot}$, due to the
initial range of the simulated distribution, and by softer drop-offs parallel
to the example TFR, due to the introduced scatter.

In the \mbox{$\dtf\!\!$~--~$\log{\vrot}$} plot, shown in panel (f), this
translates into a rectangle, the constant $\log{\vrot}$ edges are obviously
preserved, and the scatter is purely in the $\dtf$ direction.  However, the
inclusion of a maganitude cut causes a slope at the corresponding edge of the
\mbox{$\dtf\!\!$~--~$\log{\vrot}$} distribution.  For a more realistic,
magnitude-limited sample, the range in $\log{\vrot}$ would not be restricted
as it is in this example, and so the entire low $\log{\vrot}$ side of the
$\dtf$ versus $\log{\vrot}$ distribution would be sloped.  In addition, a
realistic magnitude distribution would show a cut-off at the bright end due to
the observed nature of the luminosity function.  Magnitude errors in studies
like this one do not typically exceed $0.2$ mag, and so adopting $1.0$ mag of
$M_B$ scatter would require intrinsic scatter in the TFR.  However, assuming a
uniform distribution of $\log{\vrot}$ with this amount of intrinsic scatter is
clearly at odds with the observed luminosity function.

With an unrestricted range of $\log{\vrot}$ and magnitude cuts for $M_B >
-18.64$ and $M_B < -21.63$ the distribution of points in the TFR and
\mbox{$\dtf\!\!$~--~$\log{\vrot}$} planes would look nearly identical to those
shown in panels (c) and (d).  Therefore, scatter in either $M_B$ and/or
$\log{\vrot}$, in the form of intrinsic scatter or measurement errors, when
combined with restrictions on the $M_B$ range, whether due to the underlying
luminosity function, selection effects or directly imposed, produces a
correlation in a plot of $\dtf$ versus $\log{\vrot}$.  As these restrictions
only affect the edges of the \mbox{$\dtf\!\!$~--~$\log{\vrot}$} distribution,
the strength of the correlation, and its apparent slope, are related to the
ratio of the total scatter to the $M_B$ range covered.  The
correlation is strongest, with slope equal to minus that of the TFR, when the
$M_B$ range is smaller than the scatter, and weakens, with flatter slope, as the
$M_B$ range grows relative to the scatter.

Finally, note that the combination of a scatter with restrictions on the
magnitude range leads to the measured slope of the simulated/observed TFR
appearing flatter than the real underlying TFR, as can be seen in panel (c) of
Fig.~\ref{fig:tf_res_vrot_example}.  (Likewise, restrictions in the rotation
velocity range lead to a steepening of the apparent TFR.)  However, this well
known problem is greatly reduced by choosing a method to fit the TFR which
minimises the residuals in $\log{\vrot}$, rather than $M_B$ \citep[e.g.,
see][]{Willick94}, as is done throughout this paper.

While we have made the reasonable assumption of Gaussian scatter on
$\log{\vrot}$ and $M_B$ in the above examples, the appearance of a correlation
does not depend on the form of the scatter, merely that there is some
broadening of the TFR.  Clearly the detailed distribution in the
\mbox{$\dtf\!\!$~--~$\log{\vrot}$} plane will vary with the choice of the scatter
distribution, but the existence of a significant correlation is robust.

The examples in Fig.~\ref{fig:tf_res_vrot_example} are relatively simple,
in order to avoid confusing the effects of different simulation parameters.
However, to compare more fairly, though still qualitatively, with our real
data, we require a more detailed and realistic set of simulated data.  To
generate this data we use the following method.
We assign each simulated galaxy a redshift, randomly drawn from a uniform
distribution%
\footnote{actually discrete values separated by 0.001 to improve the
  computational speed, by allowing redshift-dependent values to be calculated
  prior to the galaxy-generating loop} %
on the range 0.0--1.0.  Because of the varying volume elements enclosed by a
given solid angle at different redshifts, we must preferentially include
galaxies in our sample at a rate proportional to the comoving volume element
at their redshift.  To do this we calculate the ratio of the comoving volume
element at the assigned galaxy redshift to that at redshift 1.0.  If a random
number between 0 and 1 is greater than this ratio, then we reject this galaxy
from our sample and begin generating a new galaxy.  This assumes that the
comoving number density of galaxies is constant, which is true in the absence
of mergers, and a reasonably good assumption at these redshifts.  Changes in
the number density at a given magnitude due to luminosity evolution are
considered later.

If the galaxy is not rejected it is then assigned a true absolute magnitude,
randomly drawn from a uniform distribution between $-12.0$ and $-24.0$.  We
model the effect of the luminosity function by calculating $\phi(M_B)$,
normalised to be less than unity on the magnitude range considered, and
rejecting the galaxy from our sample if a random number between 0 and 1 is
greater than $\phi(M_B)$.  The luminosity function assumed, $\phi(M_B)$,
is a Schecter function \citep{Schechter76} with parameters from the 2dF Galaxy
Redshift Survey (2dFGRS) determined by \citet{2dfLF}, with $M^{\ast}$
converted from the $b_J$-band to the $B$-band as described in
Section~\ref{sec:phot}, i.e., $M^{\ast}=-20.3$ and $\alpha=-1.2$.

An added complication for comparing with the real data is evolution of the
luminosity function.  Without including any evolution, the bright end of the
luminosity function of the simulated galaxies does not match that seen for our
real sample.  Although our data is not a complete sample, and is subject to
complex selection effects, the bright magnitude cut-off should be reasonably
close to that of the true underlying luminosity function.  In
Section~\ref{sec:evolution} we find evidence for luminosity evolution, at a
given rotation velocity, of upto around $-1$ mag per unit redshift.
\citet{loveday04} and \citet{blanton03} find evolution in the $r$-band
luminosity function from SDSS data, although with only $z \la 0.3$, amounting to a
change in $M^{\ast}$ between $-1$ and $-2$ mag per unit redshift.  Evolution
of the luminosity function with redshift is therefore certainly plausible, and
would have most noticeable effect around the sharp cutoff at bright
magnitudes.  To account for this we apply an evolution of $M^{\ast}(z) =
M^{\ast}(0) - 1.0\,z$ when calculating $\phi(M)$ for each simulated galaxy.
This brings the luminosity function of the simulated galaxies into much better
agreement with that of our real data.

For the galaxies remaining, which now have the redshift distribution
expected given the comoving volume of space sampled at each redshift,
and an absolute magnitude distribution corresponding to a realistic,
evolving luminosity function, we now generate apparent magnitudes, and apply
a magnitude limit.

No intrinsic scatter is added to the magnitudes as we have already used a
realistic observed luminosity function to generate the magnitudes.  Adding
intrinsic scatter to these values does not make sense, as it would change the
assumed magnitude distribution to be less representative of that observed.
Any intrinsic scatter in the TFR must therefore be added to the rotation
velocities.

To apply a realistic magnitude limit we first calculate `true' apparent
magnitudes from the absolute magnitudes using the distance modulus at each
galaxy's redshift.  We then reject any galaxy with a `true' apparent magnitude
fainter than a specified limit.  We apply a magnitude limit similar to that
seen in our data, from examination of the plot of $M_B$ versus $z$ in
Fig.~\ref{fig:m_vs_z} and comparison of the faint end of the luminosity
functions for our simulated and real data.  An apparent magnitude limit of
$22.5$ mag is thus adopted.

Next, simulated `observed' $M_B$ are produced, by adding to the `true'
magnitudes values randomly drawn from a distribution with properties similar
to the measurement uncertainties on our real data.  Rather than overcomplicate
the simulation by allowing the measurement uncertainites to vary from galaxy
to galaxy, we model the magnitude measurement scatter by a Gaussian
distribution with a constant standard deviation of 0.15 mag, equal to the
median magnitude uncertainty in our real data.  Simulations with scatter based
on more realistic, variable measurement uncertainities, including using the
distribution of uncertainties from the real data, produce very similar
results, with a slight broadening of the TFR and corresponding extension of
the $\dtf$ versus $\log{\vrot}$ correlation, due to the tail of points with
large uncertainties.

We apply the scatter due to measurement errors after the magnitude cut, as the
magnitude limit is approximately due to our inability to measure the
properties of galaxies with real magnitudes fainter than the limit, rather
than a direct cut based on our observed apparent magnitudes.

`True' $\log{\vrot}$ are generated for the simulated galaxies,
using the `true' $M_B$ and the example TFR used throughout these
simulations.  These are converted into `observed' $\log{\vrot}$ by adding
random offsets to mimic intrinsic scatter and measurement errors, both drawn
from Gaussian distributions.  The measurement error standard deviation used 
was the median $\log{\vrot}$ uncertainty in our real data, $0.04$ dex.  The
intrinsic scatter standard deviation used was $0.12$ dex, corresponding to the
0.9 mag of intrinsic scatter determined in fitting our real field TFR (see
\eqnref{eqn:tf}).

The above procedure has been repeated to generate $10^6$ simulated points.
The resulting distribution of these simulated galaxies in the TFR plot is
shown by panel (a) of Fig.~\ref{fig:tf_res_vrot_sim}.  The distribution's
properties are broadly similar to those of the simpler examples in
Fig.~\ref{fig:tf_res_vrot_example} (c) and (d), but with a much more realistic
appearance.  Note that the sharp cut-off at the bright end of the luminosity
function produces a fairly well defined edge to the top of the TFR, while the
apparent magnitude limit causes a gentler drop-off at the bottom of the TFR.
This is because the wide redshift range of the sample smears out this limit in
terms of absolute magnitude.

Residuals of the simulated data points from the fiducial TFR are calculated in
the same manner as for the real data.  Panel (b) of
Fig.~\ref{fig:tf_res_vrot_sim} shows the distribution of the simulated
galaxies in the \mbox{$\dtf\!\!$~--~$\log{\vrot}$} plane.  Note the clear correlation
in this plot, despite the fact that the points were generated from a TFR
parallel to the fiducial PT92 TFR, and with only symmetrical scatter in $M_B$
and $\log{\vrot}$.  The well defined sloping top-right edge of this $\dtf\!\!$
-- $\log{\vrot}$ distribution is a consequence of the bright cut-off of the
underlying luminosity function.  Slightly less well-defined, but still clear,
is the sloped edge defining the bottom-left limit of the distribution.  This
is due to the fairly gentle cut-off at the faint end of the TFR, itself a
result of the apparent magnitude limit.  As this cut-off is gentler than at
the bright end of the TFR, the slope in the \mbox{$\dtf\!\!$~--~$\log{\vrot}$} plot
is flatter, and thus the distribution broadens slightly towards low
$\log{\vrot}$.

The \mbox{$\dtf\!\!$~--~$\log{\vrot}$} plot of our real data is shown in panel
(c) of Fig.~\ref{fig:tf_res_vrot_sim}.  These points display a correlation
very similar to that seen in the simulated data sets.  Note the similarity
with the detailed simulation in panel (b).  The only significant difference
could be argued to be the existence of a few points at lower $\log{\vrot}$ and
higher $\dtf$ than the simulated distribution.  However, remember that the
error distribution in our simulation assumes a constant measurement
uncertainty, whereas the real data shows a wide variation.  Including a
fraction of points with significantly larger measurement uncertainties causes
the \mbox{$\dtf\!\!$~--~$\log{\vrot}$} distribution to extend futher, to lower
$\log{\vrot}$ and higher $\dtf$.

The observed correlation of our data in the plot of $\dtf$ versus
$\log{\vrot}$ can therefore be easily explained as a result of the intrinsic
correlation between the two axes, combined with scatter in the TFR and
restrictions on the magnitude range, particularly due to the underlying
luminosity function.  It is therefore clear that a correlation between $\dtf$
and $\vrot$ does not necessarily imply a physical effect, and in our case it
appears to be entirely consistent with no $\vrot$-dependent change in the TFR with
redshift.

\subsection{Comparison with other studies} \label{sec:compare}

\begin{figure}
\centering
\includegraphics[height=0.45\textwidth,angle=270]%
                {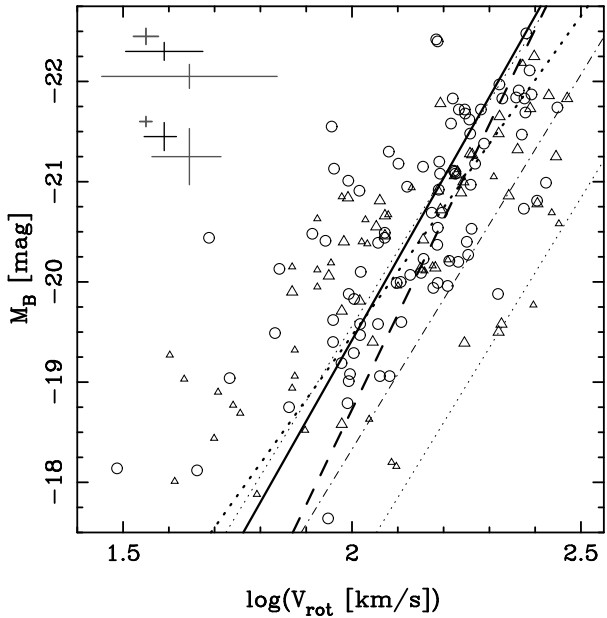}
\caption{\label{fig:bohm_tf}
        A comparison of our TFR with that of B04.  The
        circular points and solid line are the same is in Fig.~\ref{fig:tf},
        as are the thin lines indicating the fiducial PT92 TFR.
        In addition the `high quality' and `low quality' points
        from B04 are marked by large and small triangles
        respectively.  The dashed line is a fit to the `high quality'
        points, and the dotted line is a fit to the full B04
        sample.  Both of these fits are performed using the same algorithm
        used to fit our data, as described in Section~\ref{sec:TFR_fit}.
        The upper set of error bars in the top left corner indicate 
        the 10th-, 50th- and 90th-percentile uncertainties on the B04
        `high quality' points, while the lower set give the same for
        our full TFR data set.
        }
\end{figure}

The primary work with which we can compare is that of
\citet[B04]{Bohmetal04}.  This is a similar study of field spirals
between $0.1 \la z \la 1$ with $36$ `high quality' and $41$ `low
quality' rotation velocity measurements.  The internal extinction
correction applied is the same as in this paper.  In
Fig.~\ref{fig:bohm_tf} we plot our data together with those of
B04\footnote{obtained from the CDS catalogue archive}.  We also plot
TFR fits to our data and their `high quality' and full samples.  Our
fit to the B04 `high-quality' data is
\begin{equation}
M_{B} = (-9.5\pm2.3) \cdot \log{\vrot} \:+\: (0.2\pm5.1)
\end{equation}
with $\sigma_\rmn{total} = 1.2$ mag and $\sigma_\rmn{int} = 1.0$ mag,
while for their full sample we obtain
\begin{equation}
M_{B} = (-6.4\pm0.7) \cdot \log{\vrot} \:+\: (-6.7\pm2.3) \; .
\end{equation}
with $\sigma_\rmn{total} = 1.0$ mag and $\sigma_\rmn{int} = 0.8$ mag.
Our Tully--Fisher relations are thus in reasonable agreement.  This is
surprising since we see no evidence for a flattening of the TFR slope,
particularly when considering only the `high-quality' data, while B04
claim to see a strikingly shallow slope.  This may be
in some part due to their interpretation of the correlation between $\dtf$
and $\vrot$, which we have shown is potentially entirely due to correlated
errors, rather than any underlying physical effect.  They also use a
different method of fitting the TFR, a bisector fit. This may be more
easily biased to flatter slopes, by the effect of TFR scatter combined with
restrictions on the magnitude range, than our purely inverse fit (see
discussion in Section~\ref{sec:dtf_vrot}).

Apart from the disagreement over the evolution of the TFR slope, our results
are very much in accord with those of B04.  In particular we find very similar
modest evolutions in the TFR offset with redshift, of $\unisim 1$ mag by $z =
1$, also supported by the study of \citet{BLTGWF03}.

Several other earlier studies of the Tully--Fisher relation at intermediate
redshifts (e.g., \citealt{Rix97}; \citealt{SP98}) found significant luminosity
evolution with redshift, in agreement with our findings. However,
\citet*{DSS97} and \citet{Simard99} argue that much, if not all, of the
detected evolution is due to surface brightness selection effects.  This is in
agreement with the work of Vogt and her co-workers \citet{V96,V97,V02} who
find little or, more recently, no significant evolution in the TFR out to $z =
1.3$, once surface brightness selection effects have been accounted for.
Further details of this latter study's selection and analysis methods will
hopefully shed light on the difference between our results and theirs.

We will not attempt a more detailed analysis of the surface brightness
selection effects in our study given the complexity of our selection.
Nevertheless, for the above reasons and the arguments presented in
Section~\ref{sec:evolution}, we believe it is safe to take our measured
luminosity evolution as an upper limit. In Section~\ref{sec:SFR} we will use
this limit to constrain the evolution of the star-formation rate of bright
field spiral galaxies.

\section{SFR evolution with redshift} \label{sec:SFR} 
We can consider what our best-fitting evolution in the TFR implies for the
luminosity and star-formation evolution of field spirals.   The implicit
assumption in this section is that the $B$-band luminosity of a spiral galaxies
with a given rotation velocity  evolves due to its star formation history only.
For simplicity,  we shall parameterise the luminosity and star-formation rate
evolution  as simple power-laws. 

Let us assume that the rest-frame $B$-band luminosity of field spirals
evolves as a power law in $(1+z)$. Then,
\begin{equation}
L_{B}(z) = L_{B}(0)\cdot(1+z)^{\beta}
\end{equation}
\begin{equation}
\Rightarrow
\Delta M_B(z) \equiv M_{B}(z) - M_{B}(0) = - 2.5 \beta \log(1+z)
\end{equation}
In Section~\ref{sec:evolution} we find a best-fitting evolution with slope
\begin{equation}
\Delta M_{B}(z) \approx (-1.0\pm0.5) \cdot z \ \rmn{mag} \; ,
\end{equation}
which implies $\beta_\rmn{fit} = 1.3\pm0.6$.  Fitting directly with
respect to $\log(1+z)$ gives the same value for $\beta$.

Before we proceed to compare with stellar evolution models, we need a
way to relate $(1+z)$ to time.  For the concordance
cosmology,
\begin{equation}
(1+z) \approx \left(\frac{t}{t_0}\right)^{-0.813}
\end{equation}
is an excellent approximation for $z < 1$ and sufficiently accurate
for our purposes up to $z \sim 5$.  For a galaxy of age $t_\rmn{age}$,
which formed at $t_\rmn{f}$ we thus have
\begin{equation} \label{eqn:z_age}
(1+z) \approx \left(\frac{t_\rmn{f} + t_\rmn{age}}{t_0}\right)^{-0.813}
\end{equation}

Using \eqnref{eqn:z_age} to convert $(1+z)$ to $t_\rmn{age}$ for
various $t_\rmn{f}$ we can determine the expected value of $\beta$
for any star-formation history using the stellar evolution models of
\citet{BC03}.  We model the SFR as
\begin{equation}
SFR \propto \left\{
  \begin{array}{ll}
    0               &  z > z_\rmn{f}\\
    (1+z)^{\alpha} \approx 
        \left(\frac{t_\rmn{f} + t_\rmn{age}}{t_0}\right)^{-0.813\alpha} \quad &
        z \le z_\rmn{f} \; .
  \end{array}
\right.
\end{equation}
For a galaxy with constant SFR ($\alpha = 0$) and $z_\rmn{f} = 2$ we find
$\beta_{\alpha=0} = -0.27$.
This corresponds to a brightening of $0.2$ mag between $z=1$ and
today. A galaxy with constant SFR is therefore fainter in the past,
opposite to what we measure.  We therefore require a model with
a SFR which declines with time, i.e., $\alpha > 0$ 

The difference in the power-law index between our best-fitting
evolution and the constant SFR models is
\begin{equation} \label{eqn:delta_beta}
\Delta\beta_{\alpha=0} \equiv \beta_\rmn{fit} - \beta_{\alpha=0} = 1.6\pm0.6 \;.
\end{equation}
Choosing $z_\rmn{f} = 5$ only decreases this slightly to
$\Delta\beta_{\alpha=0} = 1.5\pm0.6$. We therefore work with
$z_\rmn{f} = 2$ hereafter.

In order to have some idea of of what value of $\alpha$ corresponds to
what is observed, we can assume that $B$-band luminosity is
proportional to the current SFR,
\begin{equation}  \label{eqn:lum_sfr}
\frac{L_B}{L_{B,\alpha=0}} = \frac{SFR}{SFR_{\alpha=0}}
\end{equation}
which implies
$\alpha = \Delta\beta_{\alpha=0} = 1.6\pm0.6$.
Putting this form of for the SFR (i.e., SFR $\propto (1+z)^{1.6}$)
into the \citet{BC03} code gives
\begin{equation}
\beta_{\alpha=1.6} = 1.2 \ \Rightarrow\ \Delta\beta_{\alpha=1.6} = 0.1\;,
\end{equation}
nearly but not quite zero, as \eqnref{eqn:lum_sfr} is not a perfect
approximation.  However, we can improve our estimate by a
straightforward parameter search to determine the value of $\alpha$
which produces our observed $\beta_\rmn{fit}$. We thus find $\alpha
= 1.7 \pm 1.1$.

In summary, we find that SFR$(z) \propto (1+z)^{1.7\pm1.1}$ in our sample of
relatively bright ($M_B \ga M^{\ast}_B$) galaxies. As we argued earlier, the
luminosity evolution determined from our TFR evolution is likely to be an
overestimate. The corresponding SFR evolution is therefore probably also an
upper limit, and the evolution of the average SFR in luminous spirals may be
slower than that found here. In contrast, studies of the SFR density of the
Universe at low redshift \citep[e.g.,][]{GZAR95} and as a function of
look-back time (see, e.g., \citealt{HPJD04} for a recent summary) generally
indicate a very strong evolution up to $z\sim1$. The precise rate of this
evolution is somewhat controversial, but most studies agree that,
parameterizing as SFR~$\propto (1+z)^{\alpha}$, $\alpha \simeq 3$--$4$ (see
\citealt{Hopkins04} for a compilation of SFR density evolution
measurements).

This disparity suggests that the rapid evolution in the SFR density of
the universe, observed since $z\sim1$, is \emph{not} driven by the
evolution of the SFR in individual bright spiral galaxies, like those
in our sample. However, it should be noted that, given the size of our
sample and the intrinsic scatter in the TFR, the derived constraints
on the evolution of SFR in the bright spiral galaxy population are
relatively weak.  Nevertheless, this kind of approach seems very
promising for future studies with samples of several hundred or even
thousands of galaxies to $z\sim1$.

\section{Conclusions}

We have measured confident rotation velocities for $89$ field spirals
with $0.1 \la z \la 1$, and used these to examine the evolution of the
TFR for these galaxies.  The best-fitting TFR for the full sample has
a slope which is entirely consistent with that found locally by PT92.
There is an intercept offset of $\unisim 1$ mag, such that our
galaxies are brighter for a given $\vrot$, but it is likely that this
is a least partly due to the differing methods employed to measure the
magnitudes and rotation velocities.  We therefore only evaluate
evolution in the TFR by making internal comparisons of our data.

Fitting sub-samples binned by redshift indicates that the TFR
intercept evolves by $\unisim 1$ mag between $z = 1$ and today, in the
sense that more distant galaxies are brighter for a given $\vrot$.
Plotting the residuals of our data from the local TFR against redshift
confirms this trend in TFR intercept -- higher-redshift galaxies are
offset to brighter magnitudes.  Fitting this correlation we find an
evolution of $-1.0\pm0.5$ mag by $z = 1$, which we argue is an upper
limit due to the selection effects present in our sample.

We find no significant evidence for a change in TFR slope.  Previous
studies have used an observed correlation between the TFR residuals
and $\vrot$ to argue that low mass galaxies have evolved significantly
more than those with higher mass.  However, we have demonstrated that
such a correlation may be due solely to an intrinsic coupling between
$\vrot$ scatter and TFR residuals, and thus does not necessarily
indicate a physical difference in the evolution of galaxies with
different $\vrot$.

Finally, we have used the stellar population models of \citet{BC03} to
interpret our observed TFR luminosity evolution in terms of a
star-formation rate evolution.  If the SFR in spiral galaxies had
remained constant since $z\sim1$, their $B$-band luminosity should
have increased with time. However, we find the opposite trend,
indicating that the SFR in these galaxies was larger in the past.  We
estimate SFR$(z)\propto (1+z)^{1.7\pm1.1}$ for our sample. We argue
that, given the likely selection effects, this is an upper limit on
the SFR increase with look-back time. Our results therefore suggest
that the rapid evolution in the SFR density of the universe observed
since $z\sim1$ is not driven by evolution of the SFR in individual
bright spiral galaxies.

Even though we cannot yet place strong constraints on the evolution of
the SFR in the bright spiral galaxy population, due to our relatively
small sample and the intrinsic scatter in the TFR, our approach could
be successfully applied in ongoing and future surveys.  A study with
similar quality data for a sample of $\unisim 600$ galaxies would be
needed to reduce the error in $\alpha$ to $\pm 0.4$, providing a clear
($\unisim 3 \sigma$) rejection of the hypothesis that the SFR density
of the universe and the SFR of the average spiral galaxy evolve at the
same rate.

\section*{Acknowledgments}
We would like to thank Ian Smail for generously providing additional
imaging data.  We acknowledge useful discussions with Alejandro
Garcia-Bedregal and Asmus B\"ohm concerning the PT92 zero-point. 
We also thank the anonymous referee for comments that 
greatly improved this paper. 

\bsp

{\small%

}

\label{lastpage}

\end{document}